\documentclass[journal]{IEEEtran}  

\IEEEoverridecommandlockouts
\usepackage{cite}

\usepackage{amsmath,amssymb,amsfonts}
\usepackage{algorithmic}
\usepackage{graphicx}
\usepackage{textcomp}
\usepackage{subfig}
\usepackage{xcolor}
\usepackage{booktabs}
\usepackage{varwidth}
\usepackage[T1]{fontenc}
\usepackage[font=small,labelfont=bf,tableposition=top]{caption}
\usepackage{booktabs}   
\usepackage{booktabs, makecell, tabularx, xcolor}
\usepackage{adjustbox}   
\usepackage{multirow}
\usepackage{comment}

\usepackage{cite}

\usepackage[utf8]{inputenc}

\DeclareCaptionLabelFormat{andtable}{#1~#2  \&  \tablename~\thetable}

\def\BibTeX{{\rm B\kern-.05em{\sc i\kern-.025em b}\kern-.08em
    T\kern-.1667em\lower.7ex\hbox{E}\kern-.125emX}}

\makeatletter
\newcommand{\linebreakand}{%
  \end{@IEEEauthorhalign}
  \hfill\mbox{}\par
  \mbox{}\hfill\begin{@IEEEauthorhalign}
}
\makeatother
    
\begin{document}

\bstctlcite{IEEEexample:BSTcontrol}
\title{Bayesian Optimization of Crossbar-Based Compute-In-Memory System Design for Efficient DNN Inference
}

\author{
\IEEEauthorblockN{Arnob Saha$^{1}$,
Bibhas Manna$^{2}$,
Nikhil Kotikalapudi$^{1}$,
Md Zesun Ahmed Mia$^{1}$,
Rahul Kumar$^{3}$,
Madhavan Swaminathan$^{1}$,
Abhronil Sengupta$^{1}$} \\
\IEEEauthorblockA{
$^{1}$ Penn State University, University Park, PA, USA,
$^{2}$ National Institute of Technology, Meghalaya, India,
$^{3}$ Ansys, India \\
Email: \{arnob, mvs7249, sengupta\}@psu.edu}}

\maketitle
\begin{abstract}
  Leveraging the high density and energy efficiency of Compute-In-Memory (CIM) crossbar–based Deep Neural Network (DNN) accelerators requires optimal Design Space Exploration (DSE), which becomes increasingly challenging as complex models for advanced AI workloads expand the highly non-convex design space. Among existing DSE approaches, multi-objective Bayesian Optimization (BO) is promising, as it explores high-quality design solutions while querying costly CIM simulators selectively. In this work, we propose a multi-objective BO framework that holistically co-optimizes hardware and algorithm parameters of a CIM crossbar–based hardware accelerator for various DNN inference tasks. Depending on NN model depth, our framework handles high-dimensional design spaces (with $26$ and $50$ dimensions) and extremely large search complexities on the order of $O(10^{12})$ and $O(10^{27})$ for VGG8/CIFAR-10 and VGG16/Tiny-ImageNet-200. Our method attains $91.72 \%$ and $57.2 \%$ accuracy, respectively, comparable to baseline designs, while improving chip area ($65.52 \%$ and $50.7 \%$), read latency ($9.52 \%$ and $13.27 \%$), read dynamic energy ($31.23 \%$ and $52.07 \%$) and increasing memory utilization ($13.41 \%$ and $2.67 \%$).
\end{abstract}

\vspace{-2mm}
\section{Introduction}
\begin{figure}[t]
\centering
\includegraphics[width=0.32\textwidth]{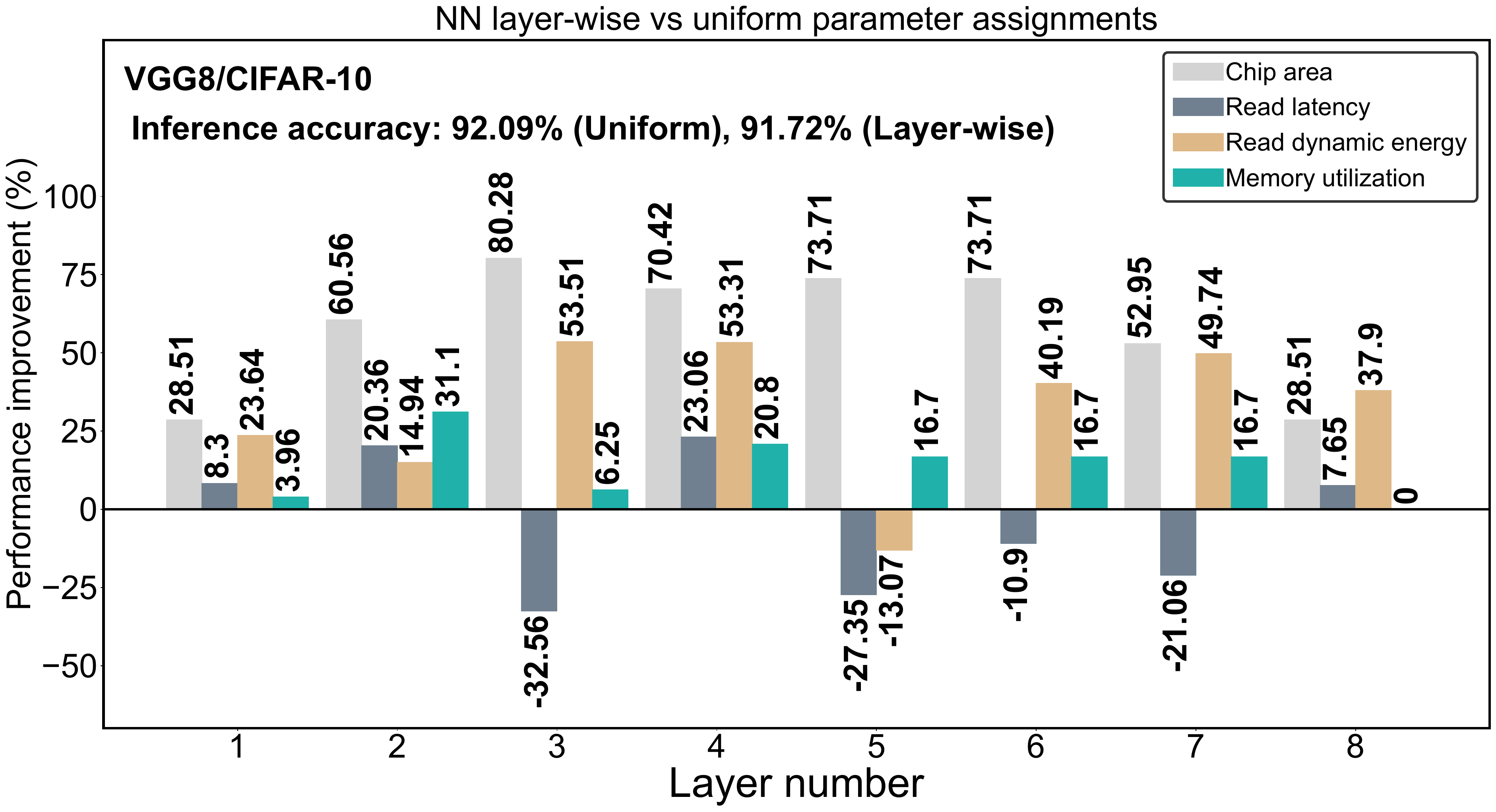}
\caption{Performance comparison between NN layer-wise and uniform parameter assignments.}
\label{layerwise}
\vspace{-5mm}
\end{figure}

Limited computation, memory, and communication resources in edge processing environments make hardware implementation of Deep Neural Networks (DNNs) challenging, as they must perform computationally expensive vector–matrix multiplications (VMMs) efficiently. In that regard, non-volatile memory (NVM) device-based CIM crossbar systems have shown superior performance over CMOS-based DNN hardware accelerators that suffer from the traditional von Neumann bottleneck \cite{jiang2020device}. 
\begin{figure}[t]
\centering
\includegraphics[width=0.33\textwidth]{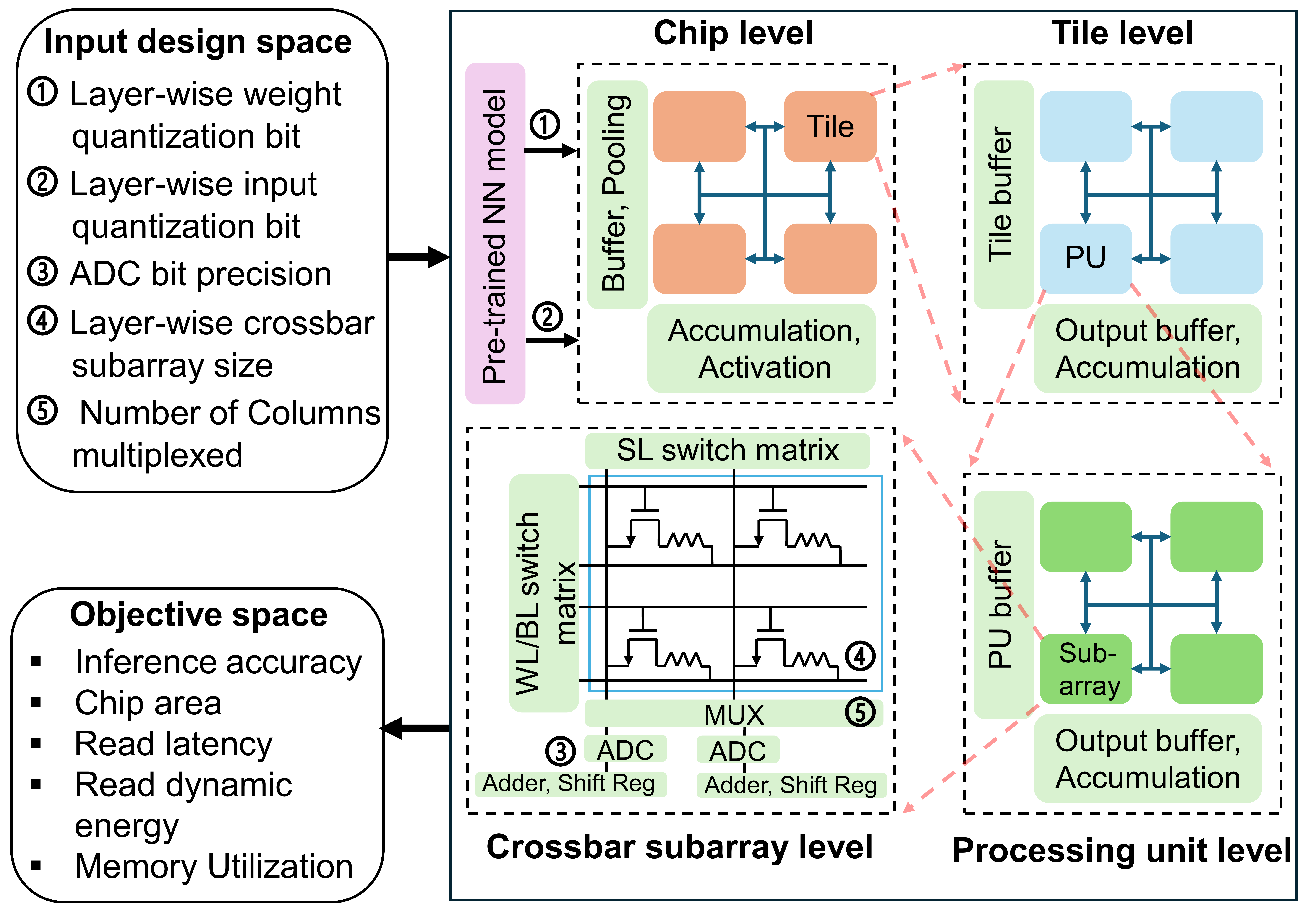}
\caption{Crossbar-based CIM system design hierarchy \cite{peng2020dnn+} and corresponding design space definitions.}
\label{heirarchy}
\vspace{-5mm}
\end{figure}
Traditionally, a CIM crossbar system is organized in a hierarchy, as shown in Fig. \ref{heirarchy}, where the bottom-most crossbar subarray level is grouped into multiple processing units, which are in turn organized into multiple tiles, and all tiles are finally integrated at the chip level. Due to the analog nature of the crossbar inputs and outputs, each level requires associated Analog-to-Digital Converters (ADCs), input/output buffers, switching matrices, and other peripheral circuits, which collectively incur significant chip area, data processing time, and energy overhead. Therefore, accurately evaluating DNN inference performance on such hardware requires careful tuning of both algorithm- and hardware-level design parameters, including weight and input quantization bits, ADC resolution, crossbar subarray size, and others. Moreover, as different NN layers can have varying workload characteristics (memory-bound or compute-bound) \cite{wang2019haq}, layer-wise parameter allocation is very useful for efficient hardware accelerator design compared to assigning identical parameter values across all NN layers that can lead to unnecessary area, latency, and energy overhead. The situation is further complicated by the fact that different DNN layers may carry varying levels of importance due to their learnt representations and therefore benefit from different design choices like weight and input bit precision \cite{chakraborty2020constructing}. Fig. \ref{layerwise} illustrates the algorithm- and hardware-level performance analysis—covering chip area, read latency, read dynamic energy, and memory utilization—for a VGG8/CIFAR-10 classification task under NN layer-wise and uniform allocations of the input design space parameters. Layer-wise parameter assignment can significantly improve overall hardware efficiency while maintaining approximately the same inference accuracy.

Nevertheless, layer-wise parameter assignment increases the dimensionality of the design space and the complexity of finding the best hardware accelerator configuration for a given NN model, especially for deeper networks and more complex datasets. In addition, different performance metrics are predominantly influenced by different design parameters, which often impose conflicting requirements. Consequently, it is crucial to obtain a single design that simultaneously improves all objectives, making the search for an optimal CIM crossbar system naturally formulated as a multi-objective optimization problem over multiple intertwined, and conflicting objectives. While there are multiple approaches to DNN hardware accelerator DSE \cite{benmeziane2021comprehensive, jiang2020device, krestinskaya2024neural, ghosh2025diffaxe, seo2025airchitect, yang2021multi}, this problem space has been relatively underexplored for memristive CIM systems, primarily due to the fact that CIM hardware simulators are computationally expensive in terms of run-time and simplified functional abstractions are not valid at scale due to continuous data representation transfer between analog and digital domains. Further, most of these approaches either assume differentiable or low-dimensional search spaces, require significant costly simulator evaluations, have slow search speed, depend on large offline datasets, or have limited generalizability. Beyond these methods, Bayesian optimization (BO) has emerged as an effective, gradient-free, sample-efficient technique for black-box optimization problems \cite{jones1998efficient}. While BO has been used for CMOS accelerator co-design \cite{shi2020using,reagen2017case}, their optimization landscape is relatively simpler than memristive CIM architectures along with much smaller search spaces. To the best of our knowledge, the feasibility, problem formulation, algorithmic workflow and effectiveness of multi-objective BO for efficiently optimizing CIM hardware accelerator designs with highly complex, high-dimensional design spaces has not yet been  explored.

\vspace{-2mm}

\section{Multi-objective Bayesian Optimization}
A multi-objective optimization problem optimizes multiple objectives jointly and is defined as $f_{mul}(x)=[f^{1}(x),f^{2}(x),\dots,f^{M}(x)]\in\mathbb{R}^{M}$, $M\geq2$, where $x$ is an input design. A solution $x$ Pareto-dominates $x'$ if $f^{m}(x)\geq f^{m}(x')$ for all $m$ and $f^{m}(x)>f^{m}(x')$ for at least one objective. All non-dominated solutions form the Pareto front, and their corresponding designs form the Pareto optimal set \cite{deb2011multiobjective}. BO solves this costly black-box optimization problem by iteratively training Gaussian Process (GP)-based surrogate models \cite{zhang2018efficient} and using acquisition functions (Lower Confidence Bound (LCB) \cite{hanaoka2021bayesian} used herein) to balance exploration and exploitation. BO is initialized by evaluating initial samples $x_{\text{init}}$ with the CIM simulator to obtain initial objective values $y_{\text{init}}$, the initial Pareto front $y_{\text{Pareto}}$, and hypervolume $hv_{\text{init}}$. Then, Gaussian surrogate models and likelihoods are constructed for each objective and jointly trained by maximizing marginal log-likelihood using Adam optimizer. At each iteration $t$, predictive mean and variance are computed for $N_{sur}$ new candidate designs, and vector-valued acquisition scores $\alpha=[\alpha_1,\alpha_2,\dots,\alpha_M]$ are used as objective proxies. Since the acquisition landscape is discrete and multi-modal, NSGA-II with Simulated Binary Crossover (SBX) and polynomial mutation optimizes the acquisition functions to obtain a non-dominated acquisition Pareto front. This front is combined with the previous Pareto front to compute the current hypervolume $hv$ and hypervolume improvement $\Delta hv$, and the design $x_{n+1}$ with maximum $\Delta hv$ is evaluated by the CIM simulator to obtain $y_{n+1}$. The surrogate models, Pareto front $Y_t$, and hypervolume $hv_t$ are updated iteratively, enabling efficient DSE of CIM systems with non-convex, discrete search spaces and costly-to-evaluate objective evaluations.

\begin{table*}[t]
\centering
\caption{Objective space performance comparison between layer-wise parameter-assigned Pareto designs and uniformly parameter-assigned designs}
\footnotesize
\setlength{\tabcolsep}{3pt}
\renewcommand{\arraystretch}{1.15}
\resizebox{\textwidth}{!}{%
\begin{tabular}{c | c | c c c c c | c c c c c}
\hline
\multicolumn{12}{c}{\textbf{Performance Comparison for VGG8/CIFAR-10 and VGG16/Tiny-ImageNet-200 Tasks}} \\
\hline
\multirow{2}{*}{\textbf{Task}} 
& \multirow{2}{*}{\textbf{Design}} 
& \multicolumn{5}{c|}{\textbf{Input design space}} 
& \multicolumn{5}{c}{\textbf{Objective space}} \\
\cline{3-12}
& 
& WBP & IBP & ABP & CSS & CCM 
& Accuracy ($\%$) & Area ($mm^{2}$) & Latency ($ms$) & Energy ($\mu J$) & Mem Util ($\%$) \\
\hline

\multirow{6}{*}{\shortstack{VGG8/\\CIFAR-10}}
& Variable WBP 
& 3/4/5 & 5 & 5 & 256 & 8 
& 91.48--92.09 & 7.1--10.5 & 0.59--0.76 & 9.55--15.18 & 62.92--83.89 \\

& Variable IBP 
& 5 & 3/4/5 & 5 & 256 & 8 
& 85.23--92.09 & 10.44--10.5 & 0.61--0.68 & 8.05--15.18 & 69.79 \\

& Variable ABP 
& 5 & 5 & 4/5 & 256 & 8 
& 80.47--92.09 & 10.4--10.5 & 0.61--0.63 & 8.97--15.18 & 69.79 \\

& Variable CSS 
& 5 & 5 & 5 & 256/128/64 & 8 
& 87.12--92.09 & 8.25--10.81 & 0.63--1.50 & 8.97--37.89 & 69.79--98.84 \\

& Variable CCM 
& 5 & 5 & 5 & 256 & 16/8/4 
& 92.09 & 10.5--10.7 & 0.52--0.92 & 14.53--16.51 & 69.79 \\

& \textbf{Layer-wise BO} 
& \textbf{3/4/5} & \textbf{3/4/5} & \textbf{4/5} & \textbf{256/128/64} & \textbf{16/8/4} 
& \textbf{91.72} & \textbf{3.62} & \textbf{0.57} & \textbf{10.44} & \textbf{83.20} \\
\hline

\multirow{6}{*}{\shortstack{VGG16/\\Tiny-ImageNet-200}}
& Variable WBP 
& 5/6/7/8 & 8 & 8 & 256 & 8 
& 55--58.2 & 206.18--306.02 & 1.67--2.32 & 109.3--173 & 90.71--94.08 \\

& Variable IBP 
& 8 & 5/6/7/8 & 8 & 256 & 8 
& 51.3--58.1 & 306.02 & 1.55--2.32 & 117.66--173 & 94.08 \\

& Variable ABP 
& 8 & 8 & 7/8 & 256 & 8 
& 58.1 & 111.01--306.02 & 2.06--2.32 & 90.43--173 & 94.08 \\

& Variable CSS 
& 8 & 8 & 8 & 256/128/64 & 8 
& 58.1 & 306.02--862.23 & 2.32--13.75 & 173--304.43 & 94.08--99.89 \\

& Variable CCM 
& 8 & 8 & 8 & 256 & 16/8/4 
& 58.1 & 153.01--612.03 & 2.16--2.66 & 171.72--175.9 & 94.08 \\

& \textbf{Layer-wise BO} 
& \textbf{5/6/7/8} & \textbf{5/6/7/8} & \textbf{7/8} & \textbf{256/128/64} & \textbf{16/8/4} 
& \textbf{57.2} & \textbf{136.61} & \textbf{1.83} & \textbf{72.3} & \textbf{95.03} \\
\hline
\end{tabular}%
}

\label{table:merged_performance_comparison}
\end{table*}

\vspace{-2mm}

\section{Results and Discussions}
\noindent \textbf{Design Specifications:} This work optimizes CIM crossbar system designs using layer-wise weight bit precision (WBP), input bit precision (IBP), and crossbar subarray size (CSS), while ADC bit precision (ABP) and crossbar columns multiplexed per ADC (CCM) are selected uniformly across layers from predefined values. The objectives are to maximize inference accuracy and memory utilization while minimizing chip area, read latency, and read dynamic energy, where read latency is the per-image processing time and memory utilization is the ratio of used memory to total memory capacity \cite{peng2020dnn+}. 

To benchmark algorithm- and hardware-level performance metrics of a crossbar-based CIM system design, NeuroSim \cite{peng2020dnn+}, a popular architecture-level simulator, has been used as the black box function. In this work, VGG8, an $8$ layer NN model, has been evaluated on CIFAR-10 dataset and VGG16, a $16$ layer NN model, has been analyzed on Tiny-Imagenet-200 dataset. The models have been pre-trained using SGD optimizer for $200$ iterations with initial learning rate of $0.01$ and standard scheduler. As shown in Fig. \ref{heirarchy}, at every BO iteration, the simulator is launched with pre-trained analog weight values and layer-wise input design space parameters to evaluate the objective space functions. For layer-wise optimization, we considered different values for each input design parameter, as listed in Table \ref{table:merged_performance_comparison}. The training dataset is used for optimization and the resulting designs are validated on the test dataset to obtain the corresponding inference accuracy.

\noindent \textbf{VGG8/CIFAR-10 Results:} Each candidate design is represented by a $26$-dimensional input vector comprising $8$ layer-wise WBP values, $8$ layer-wise IBP values, $8$ layer-wise CSS values, one ABP value, and one CCM value, while the objective space contains $5$ metrics. The resulting design-space complexity is $O(10^{12})$. The BO framework is executed for $200$ iterations, where the GP surrogate models are trained for $250$ epochs per iteration and the next design is selected from $N_{sur}=2000$ candidate samples, resulting in $116$ Pareto-optimal solutions. Since inference accuracy is the primary objective, we select a Pareto-optimal design with accuracy comparable to the best uniformly parameter-assigned baseline (note from Table \ref{table:merged_performance_comparison} that we consider several uniformly parameter-assigned designs where each input design space parameter has been varied one at a time while the remaining parameters have been kept fixed at certain values). The baseline configuration uses WBP=$5$, IBP=$5$, ABP=$5$, CSS=$256$, and CCM=$8$ uniformly across all layers, achieving $92.09\%$ accuracy, $10.5$ mm$^2$ chip area, $0.63$ ms read latency, $15.18$ $\mu$J read dynamic energy, and $69.79\%$ memory utilization. In comparison, the selected layer-wise Pareto design achieves $91.72\%$ accuracy with only $0.37\%$ degradation, while reducing area, latency, and energy by $65.52\%$, $9.52\%$, and $31.23\%$, respectively, and improving memory utilization by $13.41\%$, as reported in Table \ref{table:merged_performance_comparison}.

\noindent \textbf{VGG16/Tiny-ImageNet-200 Results:} Each design is represented by a $50$-dimensional input vector, with design-space complexity of $O(10^{27})$. Due to resource constraints, BO is run for $120$ iterations using fixed subsets of $1000$ randomly selected training samples and $1000$ randomly selected testing samples, yielding $76$ Pareto-optimal designs. The best uniform baseline uses WBP=$7$, IBP=$8$, ABP=$8$, CSS=$256$, and CCM=$8$, achieving $58.2\%$ accuracy, $277.08$ mm$^2$ area, $2.11$ ms latency, $150.84$ $\mu$J energy, and $92.36\%$ memory utilization. The selected layer-wise optimized Pareto design achieves $57.2\%$ accuracy, only $1\%$ lower than the baseline, while improving area, latency, energy, and memory utilization by $50.7\%$, $13.27\%$, $52.07\%$, and $2.67\%$, respectively, as shown in Table \ref{table:merged_performance_comparison}. 

\begin{figure}[t]

\centering
\includegraphics[width=0.35\textwidth]{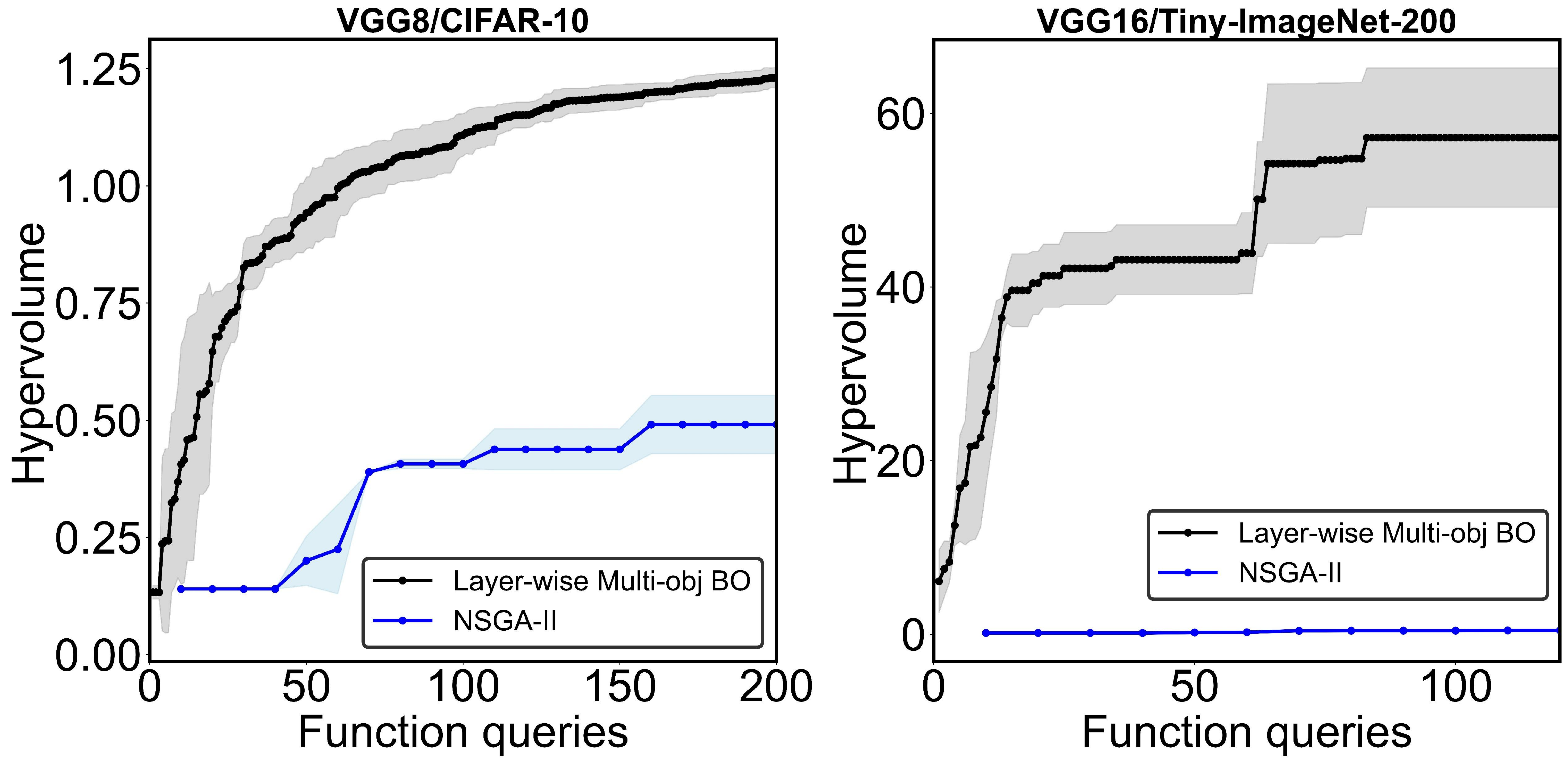}
\caption{Hypervolume comparison between multi-objective BO and NSGA-II.}
\label{hypervolume}
\vspace{-5mm}
\end{figure}

\noindent \textbf{Pareto design quality:} Pareto-front quality is evaluated using the hypervolume indicator, where higher values indicate better Pareto solutions \cite{reagen2017case}. Unlike NSGA-II \cite{deb2011multiobjective}, which requires multiple costly simulator queries per evolutionary iteration, the proposed BO framework selects only one sample per iteration to evaluate using the costly simulator, enabling more efficient optimization under an iso-computation budget. As shown in Fig. \ref{hypervolume}, BO achieves higher average hypervolume than NSGA-II across both tasks using $1000$ randomly selected training samples, with shaded regions denoting the standard deviation over $4$ independent runs.

\vspace{-2mm}
\section{Conclusions}
This work leverages the sampling intelligence of multi-objective BO and the effectiveness of layer-wise parameter allocation to efficiently design complex crossbar-based CIM systems with optimized performance metrics. 
Future work can focus on accelerating optimization convergence for more complex models and datasets by exploring alternative multi-objective BO techniques. Investigation of DSE for CIM systems catered for Transformer architectures is also highly relevant.

\vspace{-2mm}

\section*{Acknowledgments}
This material is based upon work supported primarily by the National Science Foundation under Grant No. CNS 2137259 - Center for Advanced Electronics through Machine Learning (CAEML) and its industry members. 
\vspace{-2mm}


\end{document}